\documentclass[letterpaper, 10 pt, conference]{ieeeconf}  

\IEEEoverridecommandlockouts       

\usepackage{lipsum, graphics} 
\usepackage{float}
\usepackage{algorithm}
\usepackage{algorithmic}

\usepackage{amsmath} 
\usepackage{amssymb}  
\usepackage[c2]{optidef}
\usepackage{todonotes}
\usepackage{hyperref}
\usepackage{svg}
\usepackage{subcaption}
\usepackage{float}

\usepackage{soul, xcolor}
\usepackage{cite}
\usepackage{todonotes}
\usepackage[font=small,labelfont=bf]{caption}

\title{\LARGE \bf
Multi-Agent Reach-Avoid Games: Two Attackers Versus One Defender and Mixed Integer Programming 
}

\author{ Hanyang Hu$^{*}$, Minh Bui$^{*}$ and Mo Chen
\thanks{*H. Hu and M. Bui contributed equally to this work. All authors are with the School of Computing Science, Simon Fraser University. \{hha160, buiminhb, mochen\}@sfu.ca. }%
\thanks{This work received support from the SFU-Huawei Joint Lab.}
}

\begin{document}
\setstcolor{red}

\maketitle
\thispagestyle{empty}
\pagestyle{empty}

\begin{abstract}
We propose a hybrid approach that combines Hamilton-Jacobi (HJ) reachability and mixed-integer optimization for solving a reach-avoid game with multiple attackers and defenders. The reach-avoid game is an important problem with potential applications in air traffic control and multi-agent motion planning; however, solving this game for many attackers and defenders is intractable due to the adversarial nature of the agents and the high problem dimensionality. In this paper, we first propose an HJ reachability-based method for solving the reach-avoid game in which 2 attackers are playing against 1 defender; we derive the numerically convergent optimal winning sets for the two sides in environments with obstacles. Utilizing this result and previous results for the 1 vs. 1 game, we further propose solving the general multi-agent reach-avoid game by determining the defender assignments that can maximize the number of attackers captured via a Mixed Integer Program (MIP). Our method generalizes previous state-of-the-art results and is especially useful when there are fewer defenders than attackers. We validate our theoretical results in numerical simulations. 
\end{abstract}


\section{INTRODUCTION}
The multi-agent reach-avoid game is a differential game between two adversarial teams. One team, called the attackers, aims to reach a target set of states without being captured by the opposing team while also avoiding obstacles. The opposing team, called the defenders, aims to stop as many attackers as it can from reaching the target while avoiding obstacles. The multi-agent reach-avoid game is an essential research tool, which has a wide range of applications in the area of drone control and robot logistics\cite{christensen2021roadmap, gao2018multi, roboflag1, roboflag2}.

Because of the competitive nature of the game and the high-dimensional joint state space, it is challenging to solve multi-agent reach-avoid games optimally. Some prior works have solved the game under various assumptions. In \cite{earl2007decomposition}, one team in the adversarial game is simplified to using simple motions, and the other team is assumed to be controlled under the trajectory primitive decomposition approach. In \cite{ibragimov2012multiplayer}, only one of the pursuers could be allowed to pursue one evader in each time interval with the integral constraints on control functions. In \cite{coon2017control}, both attackers and defenders are assumed to have complete knowledge of other players within a designed scope, and the attacker-defender pairs are decided based on the intersection of the isochrones. These mentioned methods work well in their respective scenarios under their assumptions. However, they sometimes cannot be generalized easily under a less restrictive set of assumptions. 

\begin{figure}[ht]
\centering
\includegraphics[width=0.8\linewidth]{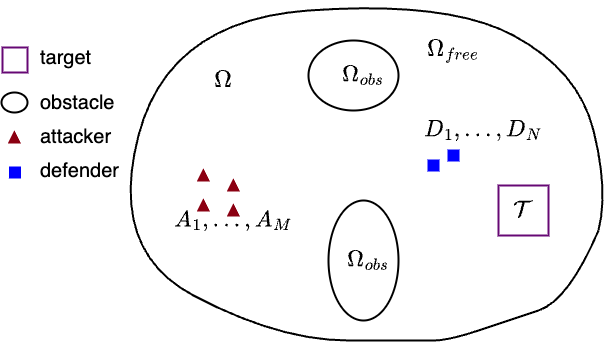}
\caption{ In a multi-agent reach-avoid game,  the attackers (red triangle) attempt to reach the target while avoiding the obstacles or being captured by the defenders along the way. The defenders (blue square) will adversarially design a strategy to stop as many attackers from reaching the target. }
\label{fig: formulation}
\end{figure}
Some researchers try to find analytical solutions to the multi-agent reach-avoid game based on geometric features. In \cite{chen2022reach}, the authors introduce a geometric method that considers the geometric relation between the target region and the reaching region in a 2 vs. 1 reach-avoid game. In \cite{8279644, yan2021cooperative}, Rui et al propose different analytical solutions based on disparate geometric features of a 2 vs. 1 game. In \cite{9067037}, the authors propose three different control strategies based on three different stages between the single attacker and the single defender. These methods could obtain suboptimal or optimal analytical solutions through case-by-case discussions. However, these methods depend on the geometry of the games and can have difficulty generalizing to different environments.

Hamilton-Jacobi (HJ) reachability analysis is a powerful method that can provide general optimal strategies for both attackers and defenders in a multi-agent reach-avoid game \cite{chen2018hamilton, mitchell2005time}. The game result only depends on the initial joint states of all attackers and defenders, and optimal joint state feedback strategies can be derived for both teams. However, HJ reachability analysis scales poorly with the problem dimensionality, restricting the previous application of the method to only the 1 vs. 1 game, a 4-dimensional problem. By leveraging the recently developed OptimizedDP library \cite{bui2022optimizeddp}, computations up to 7 dimensions become possible, allowing the numerically convergent optimal solution to be computed for games involving 3 agents (with single integrator dynamics, a common assumption in the reach-avoid game literature). However, despite these advances, when the dimensionality of the joint state of all players is greater than 6, the multi-agent reach-avoid game again becomes computationally intractable. 

Methods that use high-level logic to scale up low-dimensional solutions of HJ reachability have been effective in obtaining approximate solutions to high-dimensional problems. In \cite{chen2014multiplayer, chen2014path, Chen17, yan2022matching}, the authors propose using the maximum matching method to extend the result of a 1 vs. 1 reach-avoid game to a multi-agent case through a maximum matching approach. In \cite{chen2016multi}, a MIP is proposed to decide the assignment logic between pairs of vehicles when there are more than two vehicles to guarantee safety. These methods are based on the result of a 1 vs. 1 pursuit-evasion game and could be improved by combining more tractable HJ reachability results. 

In this paper, we propose to combine recent computational advances of high-dimensional HJ reachability analysis using OptimizedDP toolbox \cite{bui2022optimizeddp} with a new MIP that provides high-level logic to solve and scale up the solutions of multi-agent reach-avoid games. Specifically, our contributions are as follows. First, we propose a formulation based on HJ reachability for optimally solving a reach-avoid game involving 2 attackers playing against 1 defender. Our solution involves computing a reach-avoid set in a 6-dimensional state space, enabled by OptimizedDP \cite{bui2022optimizeddp} as well as the reach-avoid set for a 1 vs. 1 game. Given the 1 vs. 1 and 2 vs. 1 game solutions, we propose a MIP that  maximizes the number of attackers that are guaranteed to be captured by the defenders. The MIP allows a single defender to be assigned to defend against up to 2 attackers. We demonstrate and validate our theoretical contributions in numerical simulations of various game configurations.

The rest of this paper is organized as follows. In Section \ref{sec2}, the multi-agent reach-avoid game is formulated along with the player dynamics. In Section \ref{sec3}, we introduce the HJ reachability analysis method for solving reach-avoid games and an optimization program based on a MIP that maximizes the number of captured attackers. In Section \ref{sec4}, we show the performance of our proposed method in simulation with different configurations. In Section \ref{sec5}, we draw conclusions and discuss possible future work.
\section{PROBLEM FORMULATION}
\label{sec2}

In the multi-agent reach-avoid game, we consider $M$ attackers, $\{ A_i \}_{i=1}^M = \{ A_1, A_2,..., A_M\}$ and $N$ defenders, $\{ D_i \}_{i=1}^N = \{ D_1, D_2,..., D_N\}$. Both attackers and defenders are confined to some compact domain $\Omega \subset \mathbb{R}^2$.
The domain $\Omega$ is divided into two subsets: $\Omega_{free}$ and $\Omega_{obs}$.  Each player can move freely in $\Omega_{free}$ but should not enter $\Omega_{obs}$. We assume the dynamics of attackers and defenders are as follows:
\begin{equation}
\begin{array}{lll}
\dot{\boldsymbol{x}}_{A_i}(t)=v_A \boldsymbol{a}_i(t), & \boldsymbol{x}_{A_i}(0)= \boldsymbol{x}_{A_i}^0, & i = 1, 2, ..., M \\
\dot{\boldsymbol{x}}_{D_j}(t)=v_D \boldsymbol{d}_j(t), & \boldsymbol{x}_{D_j}(0)= \boldsymbol{x}_{D_j}^0, & j = 1, 2, ..., N
\end{array}
\end{equation}
where $\boldsymbol{x}_{A_i}$ and $\boldsymbol{x}_{D_j}$ are the 2-dimensional states, representing positions of attackers and defenders respectively, $\boldsymbol{x}_{A_i}(0)$ and $\boldsymbol{x}_{D_j}(0)$ are the initial positions of attackers and defenders respectively, $v_A$ and $v_D$ are the maximum speed of attackers and defenders respectively, $\boldsymbol{a}_i(t)$ and $\boldsymbol{d}_j(t)$ are the control inputs of attacker $i$ and defender $j$ respectively at the time $t$. 

In the reach-avoid game, the attackers aim to reach some target $\mathcal{T} \subset \Omega_{free}$, which is a compact subset of the free region, while avoiding obstacles and being captured by defenders. On the other hand, the defenders will try to prevent the attackers from reaching the target by capturing them or delaying them indefinitely. An attacker $A_i$ is considered to be captured by the defender $D_j$ when their relative distance is within a capture distance $R_C$. With that, we define the capture sets $\mathcal{C}_{ij} \subset \Omega_{free}$ for each attacker-defender pair $(A_i, D_j)$, $i = 1, ...,M, j = 1,..., N$ as $\mathcal{C}_{i j}=\left\{(\boldsymbol{x}_{A_i}, \boldsymbol{x}_{D_j}) \mid \left\|\boldsymbol{x}_{A_i}-\boldsymbol{x}_{D_j}\right\|_2 \leq R_C\right\}$. 

In multi-agent reach-avoid game, we are most interested in the number of attackers that the defenders are guaranteed to be able to prevent from reaching the target. An illustration of our multiplayer reach-avoid game is shown in Fig. \ref{fig: formulation}.
\section{METHODOLOGY}
\label{sec3}
HJ reachability analysis is a numerically convergent framework for obtaining optimal solutions to differential games \cite{fisac2015reach, chen2018hamilton, huang2011differential, mitchell2007toolbox}. We will first introduce the general background of HJ reachability and how it is used to generate the closed-loop feedback control inputs. Then, we describe the details of solving a 1 vs. 1 and a 2 vs. 1 reach-avoid game using HJ reachability. Finally, we will propose a MIP to solve the general M vs. N reach-avoid game utilizing the solutions of the 1 vs. 1 and the 2 vs. 1 game.

\subsection{Hamilton-Jacobi Reachability}
In HJ reachability analysis, one models the dynamics of a system using the following set of differential equations:
\begin{equation}
\label{dynamics}
\dot{\mathbf{x}}=f(\mathbf{x}, \mathbf{u}, \mathbf{d}), \mathbf{x}(0)=\mathbf{x}^{\mathbf{0}}
\end{equation}

\noindent where $\mathbf{x} \in \mathbb{R}^n$,  $\mathbf{u} \in \mathcal{U}$ is the joint control input of all attackers, $\mathbf{d} \in \mathcal{D}$ is the joint control input of all defenders. The sets $\mathcal{U}$ and $\mathcal{D}$ are the sets of measurable control functions for attackers and defenders respectively. In the case of the reach-avoid game we are considering, we have the following: $n = 2(M+N)$ is the joint states of $M$ attackers and $N$ defenders.

In the reach-avoid game, we need to specify the target set $R$ which is the winning condition of the attackers. Besides, we also need to specify the constraint which we should obey during the backward propagation from the target set $R$. The constraint is described by the avoid set $A$. The details of these two sets are in Section \ref{MethodologyB} and Section \ref{MethodologyC}. 

Specifically, given the target set $R$, the level set representation of $R$ is the function $l(x): \mathbb{R}^n \rightarrow \mathbb{R}$ such that $R =\left\{\mathbf{x} \in \mathbb{R}^n \mid l(\mathbf{x}) \leq 0\right\}$. Similarly, the level set representation of the avoid set $A$ is the function $g(x): \mathbb{R}^n \rightarrow \mathbb{R}$ such that $A =\left\{\mathbf{x} \in \mathbb{R}^n \mid g(\mathbf{x}) > 0\right\}$. Based on these notations, let the value function $\Phi: \mathbb{R}^n \times[-T, 0] \rightarrow \mathbb{R}$ be the viscosity solution to the Hamilton-Jacobi-Isaacs (HJI) variational inequality \cite{fisac2015reach}:
\begin{equation}
\label{HJI}
\begin{aligned}
\max \{  & \min \left\{\frac{\partial \Phi}{\partial t} + H \left(\mathbf{x}, \frac{\partial \Phi}{\partial \mathbf{x}}\right), l(\mathbf{x})-\Phi(\mathbf{x}, t) \right\}, \\
& \, \left.g(\mathbf{x})-\Phi(\mathbf{x}, t)\right\}=0, \quad t \in[-T, 0], \mathbf{x} \in \mathbb{R}^n,
\end{aligned}
\end{equation}
where the optimal Hamiltonian $H$ is calculated as:
\begin{equation*}
H(\mathbf{x}, p)=\min _{u \in \mathbb{U}} \max _{d \in \mathbb{D}} p^T f(\mathbf{x}, \mathbf{u}, \mathbf{d})
\end{equation*}
where $p = \frac{\partial \Phi}{\partial \mathbf{x}}$.

Analytical solutions to Eq. \ref{HJI} are rarely feasible, and typically solved using numerical methods \cite{OsherBook}, which depends on the discretion of the state space and dynamic programming iteration; therefore, the computational and space complexity grows exponentially as the number of states increases. Software toolbox such as OptimizedDP \cite{bui2022optimizeddp}, utilizes modern computing power to solve Eq. \ref{HJI} for up to $n=7$ with decent grid resolutions. 

Once the solution $\Phi$ has been obtained, the set of initial positions where attackers are guaranteed to win within the time interval $T$ is as follows \cite{mitchell2002application, mitchell2005time}:
\begin{equation}
\mathcal{R} \mathcal{A}_T(R, A):=\left\{\mathbf{x} \in \mathbb{R}^n \mid \Phi(\mathbf{x},-T) \leq 0\right\}
\end{equation}

In reachability literature, such set obtained from Eq. \ref{HJI} is called reach-avoid tube. Next, the optimal controls for all attackers and defenders can be derived from the solution $\Phi$ respectively \cite{huang2011differential}:
\begin{equation}
\begin{aligned}
& \mathbf{u}^*(\mathbf{x}, t)=\arg \min _{\mathbf{u} \in \mathbb{U}} \max _{\mathbf{d} \in \mathbb{D}} p(\mathbf{x},-t)^\top f(\mathbf{x}, \mathbf{u}, \mathbf{d}), t \in[0, T] \\
& \mathbf{d}^*(\mathbf{x}, t)=\arg \max _{\mathbf{d} \in \mathbb{D}} p(\mathbf{x},-t)^\top f\left(\mathbf{x}, \mathbf{u}^*, \mathbf{d}\right), t \in[0, T]
\end{aligned}
\end{equation}
where $p = \frac{\partial \Phi}{\partial \mathbf{x}}$. Furthermore, if we take the time interval $T \rightarrow \infty$, we will obtain the set of initial positions where attackers are guaranteed to win which is denoted as $\mathcal{R} \mathcal{A}_{\infty}(R, A)$. A more in-depth introduction to HJ reachability can be found in \cite{bansal2017hamilton, chen2018hamilton}.

For practical problems that involve multiple defenders and attackers (where n > 7), it is intractable to obtain $\Phi$ numerically. In the next few sections, we will describe how we solve Eq. \ref{HJI} for small sub-problems, which are then used to approximate a winning control strategy for the defenders in larger problem instances. In this paper, we will shortly denote $1$ vs. $1$ to refer to a game where one attacker is against one defender and $2$ vs. $1$ for a game where two attackers are up against one defender.

\subsection{The 1 vs. 1 Reach-Avoid Game (n = 4)}
\label{MethodologyB}

In a 1 vs. 1 reach-avoid game, the attacker wins if the attacker can reach the target set and loses the game if it is captured by the defender. Following the work in \cite{chen2014multiplayer}, the target set $R^{11}$ and the avoid set $A^{11}$ for a 1 vs. 1 reach-avoid game are defined as follows:
\begin{equation}
\label{RA_11}
\begin{aligned}
R^{11}= & \left\{\mathbf{x} \in \Omega^2 \mid \boldsymbol{x}_{A} \in \mathcal{T} \wedge\left\|\boldsymbol{x}_{A}-\boldsymbol{x}_D\right\|_2>R_C\right\} \\
& \cup \left\{\mathbf{x} \in \Omega^2 \mid x_D \in \Omega_{o b s}\right\} \\
A^{11} = & \left\{\mathbf{x} \in \Omega^2 \mid\left\|\boldsymbol{x}_{A}- \boldsymbol{x}_D\right\|_2 \leq R_C\right\} \\
& \cup\left\{\mathbf{x} \in \Omega^2 \mid \boldsymbol{x}_{A} \in \Omega_{o b s}\right\}
\end{aligned}
\end{equation}

Given these sets, we can solve Eq. \ref{HJI} to obtain the 4-dimensional reach-avoid tube $\mathcal{RA}^{11}_{\infty}(R^{11}, A^{11})$ for the 1 vs. 1 game. An attacker wins if and only if the initial joint state $(\boldsymbol{x}_{A}^0, \boldsymbol{x}_{D}^0)$ lies within the reach-avoid tube $\mathcal{RA}^{11}_{\infty}(R^{11}, A^{11})$ under the optimal control \cite{chen2014multiplayer}:
\begin{equation}
\boldsymbol{a}^*\left(\boldsymbol{x}_A, \boldsymbol{x}_D, t\right)=-v_A \frac{p_a\left(\boldsymbol{x}_A, \boldsymbol{x}_D,-t\right)}{\left\|p_a\left(\boldsymbol{x}_A, \boldsymbol{x}_D,-t\right)\right\|_2} 
\end{equation}

\noindent where $p = (p_{a}, p_{d}) = \frac{\partial \Phi^{11}}{\partial (\boldsymbol{x}_{A}, \boldsymbol{x}_{D})}$ with $p_a,p_d\in\mathbb R^2$ representing components of the gradient corresponding to the states of the attacker and defender, respectively.

If the initial joint state $(\boldsymbol{x}_{A}^0, \boldsymbol{x}_{D}^0)$ lies out of the reach-avoid tube $\mathcal{RA}^{11}_{\infty}(R^{11}, A^{11})$, the defender will win the game with the optimal control \cite{chen2014multiplayer}:
\begin{equation}
\label{1v1d}
\boldsymbol{d}^*\left(\boldsymbol{x}_A, \boldsymbol{x}_D, t\right)=v_D \frac{p_d\left(\boldsymbol{x}_A, \boldsymbol{x}_D,-t\right)}{\left\|p_d\left(\boldsymbol{x}_A, \boldsymbol{x}_D,-t\right)\right\|_2} 
\end{equation}

\subsection{The 2 vs. 1 Reach-Avoid Game (n = 6)}
\label{MethodologyC}

With the recent advances in solving high-dimensional HJ PDE \cite{bansal2021deepreach,bui2022optimizeddp,li2022infinite}, we are not limited to just solving a 1 vs. 1 reach-avoid game (by solving a 4-dimensional HJI PDE)
\cite{chen2014multiplayer}. In this paper, we will utilize the OptimizedDP toolbox \cite{bui2022optimizeddp} to obtain optimal solutions to the 2 vs. 1 game, in which one defender can be playing against up to 2 attackers in a 6-dimensional joint state space. In this 2 vs. 1 game, a defender wins the game if it is guaranteed to capture and/or stop both attackers from reaching the target set.

In the 2 vs. 1 game, the joint state is $\mathbf{x} = (\boldsymbol{x}_{A_1}, \boldsymbol{x}_{A_2}, \boldsymbol{x}_{D})$. Since we are interested in guaranteeing that the defender can stop both attackers and win the game, we will consider a 2 vs. 1 reach-avoid game that computes the region of states where \textbf{at least} one of the attackers is guaranteed to reach the target, the complement of this region is a set of states where none of the attackers can arrive at the target or the defender is guaranteed to win.

First, we define the target set $R^{21}$ and the avoid set $A^{21}$ as the following:
\begin{equation}
\label{RA_21+}
\begin{aligned}
R^{21}&=  \left\{\mathbf{x} \in \Omega^3 \mid \boldsymbol{x}_{A_1} \in \mathcal{T} \wedge\left\|\boldsymbol{x}_{A_1}- \boldsymbol{x}_D\right\|_2>R_C \}\right. \\
& \left. \cup \quad \{\mathbf{x} \in \Omega^3 \mid \boldsymbol{x}_{A_2} \in \mathcal{T} \wedge\left\| \boldsymbol{x}_{A_2}- \boldsymbol{x}_D\right\|_2>R_C\right\} \\
& \cup\left\{\mathbf{x} \in \Omega^3 \mid \boldsymbol{x}_D \in \Omega_{o b s}\right\} \\
A^{21}&=  L_{1} \cup L_{2}\\
\end{aligned}
\end{equation}
where
\begin{equation}
\label{L}
\begin{aligned}
L_{1} = &\left(\{\mathbf{x} \in \Omega^3 \mid\left\| \boldsymbol{x}_{A_1}- \boldsymbol{x}_D\right\|_2 \leq R_C\right\} \\ 
& \cup\left\{\mathbf{x} \in \Omega^3 \mid \boldsymbol{x}_{A_1} \in \Omega_{o b s}\}\right)\\
&\cap \{\mathbf{x} \in \Omega^3 \mid (\boldsymbol{x}_{A_2}, \boldsymbol{x}_{D}) \notin \mathcal{RA}^{11}_{\infty} \}\\\
L_2 = &\left( \left\{\mathbf{x} \in \Omega^3 \mid\left\| \boldsymbol{x}_{A_2}- \boldsymbol{x}_D\right\|_2 \leq R_C\right\}\right. \\
& \cup\left\{\mathbf{x} \in \Omega^3 \mid \boldsymbol{x}_{A_2} \in \Omega_{o b s}\}\right) \\
& \cap \{\mathbf{x} \in \Omega^3 \mid (\boldsymbol{x}_{A_1}, \boldsymbol{x}_{A_D}) \notin \mathcal{RA}^{11}_{\infty}  \}\\
\end{aligned}
\end{equation}

In the above definition, $R^{21}$ comprises states where at least one attacker arrives at the target set $\mathcal{T}$ while not being captured by the defender, or the defender is obstructed by  $\Omega_{obs}$. The avoid set $A^{21}$ is the union of $L_1$ and $L_2$. $L_1$ is a set of states where the attacker $A_1$ has been captured or hit obstacles and the attacker $A_2$ be captured in the future time steps, which happens when $(\boldsymbol{x}_{A_2}, \boldsymbol{x}_{D}) \notin \mathcal{RA}^{11}_{\infty}$. Similarly, $L_2$ is a set of states where $A_2$ has been captured and $A_1$ is captured in future time steps. The avoid set $A^{21}$ is then the union of $L_1$ and $L_2$, which cover all the states leading both attackers getting captured. Note that the definition of $L_1$ and $L_2$ involve the 4-dimensional reach-avoid tube $\mathcal{RA}^{11}_{\infty}$, which allows us to circumvent complications involving defining a time-varying avoid set that encodes the defender capturing the attackers not at the same time, but one after another.

By solving the Eq. \ref{HJI}, we can obtain the reach-avoid tube $\mathcal{R} \mathcal{A}_{\infty}^{21}$, which is a 6-dimensional value function that represents states where at least one attacker can reach the target safely without being captured. In this case, if the initial state does not lie in the reach-avoid tube $\mathbf{x}^{0} \notin \mathcal{R} \mathcal{A}_{\infty}^{21}$, then the defender is guaranteed to capture both attackers successfully using the optimal controls:
\begin{equation}
\label{2v1d}
\boldsymbol{d}^* \left(\boldsymbol{x}_{A_1}, \boldsymbol{x}_{A_2}, \boldsymbol{x}_{D}, t\right) = v_D \frac{p_{d}\left(\boldsymbol{x}_{A_1}, \boldsymbol{x}_{A_2}, \boldsymbol{x}_{D},-t\right)}{\left\|p_{d}\left(\boldsymbol{x}_{A_1}, \boldsymbol{x}_{A_2}, \boldsymbol{x}_{D},-t\right)\right\|_2}
\end{equation}
where $p = (p_{a1}, p_{a2}, p_{d}) = \frac{\partial \Phi^{21}}{\partial (\boldsymbol{x}_{A_1}, \boldsymbol{x}_{A_2}, \boldsymbol{x}_{D})}$ with $p_{a1}, p_{a2},p_d\in\mathbb R^2$ representing components of the gradient corresponding to the states of the first attacker, second attacker, and defender, respectively.

\subsection{ Mixed Integer Programming for Assigning Attackers}
\label{sec3.3}

HJ reachability analysis is computationally intractable when the total number of attackers and defenders is greater than three. Hence, to utilize the results of the 1 vs. 1 and 2 vs. 1 game via HJ reachability, we propose a MIP that maximizes the number of attackers that can be assigned to a defender, which can be conveniently solved by software library such as the Python-MIP \cite{santos2020mixed, gurobi, cplex2009v12} Mathematically, this optimization is defined as the following:
\begin{maxi!}
{}{ \sum\limits_{i, j} e_{i j}} 
{}{}
\addConstraint{e_{i j}}{\in\{0,1\} \quad \forall i, j} \label{c1}
\addConstraint{\sum\limits_{i} e_{i j}}{\leq 2, \quad  \forall j} \label{c2}
\addConstraint{\sum\limits_{j} e_{i j}}{\leq 1, \quad  \forall i} \label{c3}
\addConstraint{e_{sj}}{= 0, \quad \forall j, \forall s \in \mathcal{I}_{j}} \label{c4}
\addConstraint{e_{kj} + e_{lj}}{\leq 1, \quad \forall j, \forall (k,l) \in \mathcal{P}_{j}} \label{c5}
\end{maxi!}

In this program, we denote the capture relationship between the attacker $A_i$ and the defender $D_j$ as $e_{i j} \in \{0,1\}$. When $e_{i j} = 1$, it indicates that the defender $D_j$ is assigned to capture the attacker $A_i$ and $e_{ij} = 0$ otherwise. 

We will then construct the constraints for this optimization program. Firstly, the constraint (\ref{c2}) limits the total number of attackers assigned to defender $D_j$ to be at most two at any time. This upper bound is limited by the computational capability for HJ reachability analysis. Secondly, the constraint (\ref{c3}) ensures that one attacker can only be captured by at most one defender to avoid the case where multiple defenders are assigned to only one attacker.

Let $\mathcal{I}_j$ be the set of all attackers that defeat the defender $\mathcal{D}_j$ in a 1 vs. 1 game when using optimal control. The constraint (\ref{c4}) makes sure that a defender $D_j$ cannot be assigned to an attacker $i$ that it cannot win in a 1 vs. 1 game by directly setting $e_{ij} = 0$. 
Using a pre-computed reach-avoid tube $\mathcal{RA}^{11}_{\infty}$, the set $\mathcal{I}_j$ can easily be constructed.

Similarly, using a pre-computed reach-avoid tube $\mathcal{R} \mathcal{A}_{\infty}^{21}$, we can construct the set $\mathcal{P}_{j} = \{ (k, l) \mid (\mathbf{x}_{A_k}, \mathbf{x}_{A_l}, \mathbf{x}_{D_j}) \in \mathcal{R}\mathcal{A}_{\infty}^{21},  0 \leq k,l \leq M, k \neq l $ \} that contains all the \textbf{distinct} pairs of attackers that can win against defender $D_j$ in a 2 vs. 1 game. 
The constraint (\ref{c5}) ensures that defender $D_j$ will not be assigned to defend against 2 attackers that $D_j$ cannot capture both.

\subsection{Iterative Capture Algorithm}
The solution of our MIP in Section \ref{sec3.3} determines the capturing assignment for each defender based on the current states of all agents. As the states of these agents will change over time, the constraints \ref{c4} and \ref{c5} of our MIP also need to be updated accordingly. Hence, we set up and solve the MIP iteratively after every time step to take into account these changes. Our iterative capture algorithm can be summarized as shown in the Algorithm \ref{alg1}.

\begin{algorithm}
\caption{Iterative Capture Algorithm}
\label{alg1}
\small
\begin{algorithmic}
\REQUIRE $\mathcal{R} \mathcal{A}_{\infty}^{11}$, $\mathcal{R} \mathcal{A}_{\infty}^{21}$
\renewcommand{\algorithmicensure}{\textbf{Initialization:}}
\ENSURE states of all attackers and defenders
\WHILE{not all attackers captured or arrive at target}
\STATE determine the constraint set  $\mathcal{I}_{j}$ for each defender $D_j$
\STATE determine the constraint set $\mathcal{P}_{j}$ for each defender $D_j$
\STATE solve the MIP
\STATE apply optimal controls for defenders based on Eq.\eqref{1v1d}, Eq.\eqref{2v1d} and the MIP's assignment result
\STATE update states of all attackers and defenders after $\Delta t$ 
\ENDWHILE
\end{algorithmic}
\end{algorithm}

Due to the optimality of the HJ controller, the defender assigned to attackers is guaranteed to capture the attackers using the optimal control inputs regardless of what control the attackers use. Hence, the above capture algorithm has this property: the guaranteed maximum number of attackers that can reach the target never increases; instead, it can decrease depending on the trajectories of the players as the game plays out and new attacker assignments are made. 
\section{NUMERICAL SIMULATIONS}
\label{sec4}

All HJ reach-avoid tubes are computed offline using the OptimizedDP library developed at Simon Fraser University \cite{bui2022optimizeddp}. First, we calculate the 1 vs. 1 and 2 vs. 1 reach-avoid tube $\mathcal{RA}^{11}_{\infty}$ and $\mathcal{R} \mathcal{A}_{\infty}^{21}$ respectively. The computation of $\mathcal{RA}^{11}_{\infty}$ is done on a $45 \times 45 \times 45 \times 45$ 4-dimensional grid, which takes a few seconds to compute until convergence and consumes 100 megabytes of RAM. And the set $\mathcal{R} \mathcal{A}_{\infty}^{21}$ is computed on a $30 \times 30 \times 30 \times 30 \times 30 \times 30$ 6-dimensional grid, which takes around 1 hour and 30 minutes to compute until convergence, and consumes around 20 gigabytes of RAM on a 16-thread Intel(R) Core(TM) i9-9900K CPU at 3.60GHz.  

We verify the correctness of our reachability calculation and the effectiveness of our MIP method in the following simulations. In our experiment, the attackers use a suboptimal control strategy that takes the shortest path to the target $\mathcal{T}$ while avoiding obstacles. For comparison, the method that uses maximum matching to obtain approximate solutions to the multi-agent reach-avoid game in \cite{chen2014multiplayer} is chosen as the baseline. In this baseline method, the maximum number of attackers matched in every time step is at most the same as the number of defenders.

To show the advantages of our method, the number of defenders is chosen to be smaller than the number of attackers in the following simulations. In the simulations, the speed of defenders and attackers are set as $v_D = 1.5$ unit per second and $v_A = 1.0$ unit per second respectively. They are all confined to a square domain with two obstacles and a target set $\mathcal{T}$ as shown in Fig. \ref{2v1_RAS}. The capture distance $R_C$ is chosen as $0.1$ units.

Once an attacker has been captured, it will be removed from the game and marked by a plus sign in the figures while the defender's past trajectory will be erased. To clearly visualize the trajectory of multiple players throughout the game, we plot the trajectories of defenders every eight-time step and remove the overlapping parts with other players. In each figure, the green edges connecting a defender to an attacker represent the capturing assignment results from our MIP and the maximum matching. 

\begin{table}[h!]
\centering
\begin{tabular}{|llll|}
\hline
\multicolumn{4}{|c|}{\textbf{Attackers Captured}} \\ \hline Game config.
           & 2 vs. 1    & 6 vs. 2    & 8 vs. 4    \\ \hline
Baseline \cite{chen2014multiplayer}   & 1          & 4          & 6          \\
Our Method & \textbf{2} & \textbf{6} & \textbf{8} \\ \hline
\end{tabular}
\caption{Number of attackers captured for different game scenarios. It can be observed that our method results in better capture rates compared to the baseline in all the games. }
\label{table:1}
\end{table}

\subsection{Reach-Avoid Tube Visualization}

\begin{figure}[ht]
\centering
\includegraphics[width=1.0\linewidth]{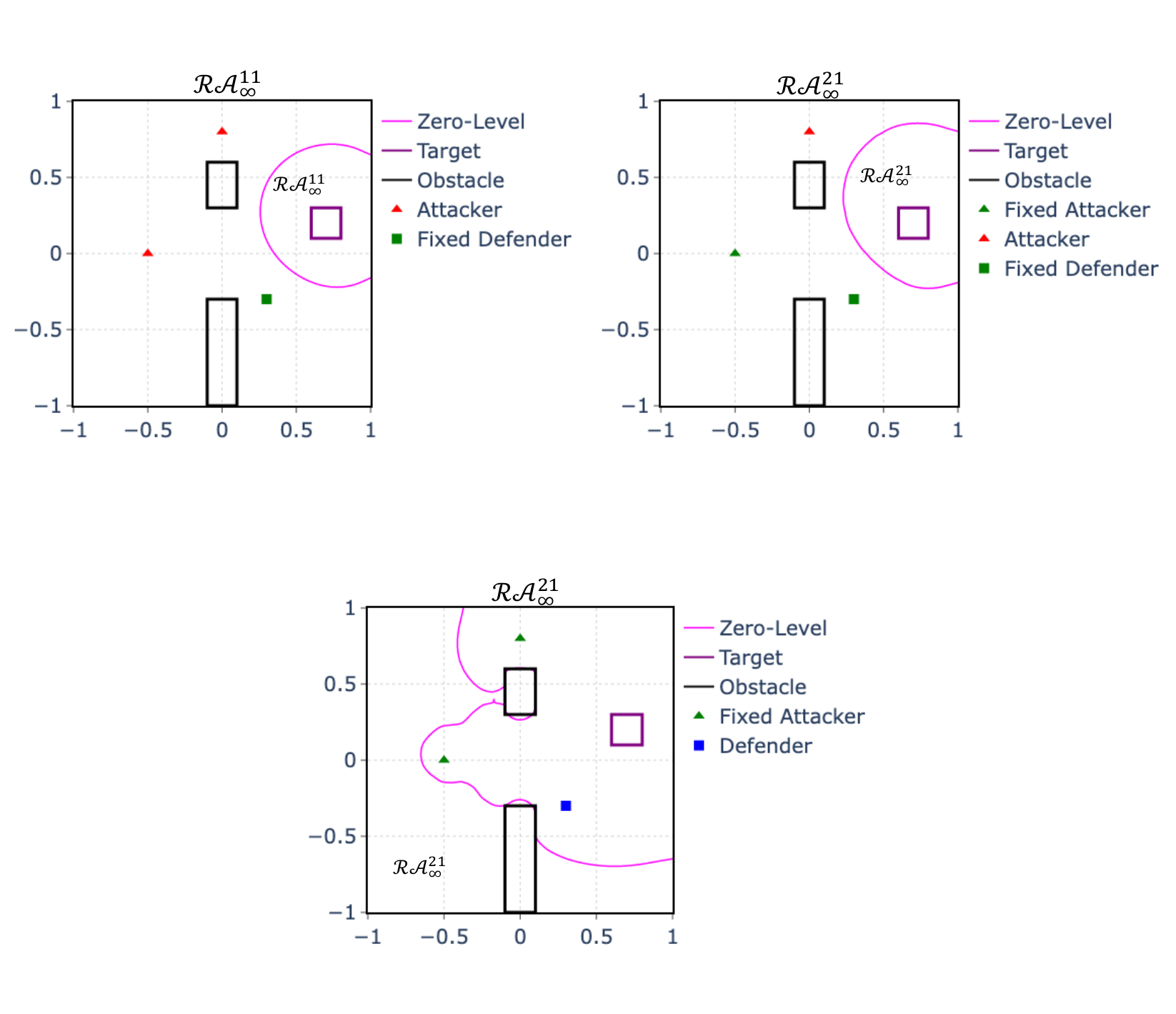}
\caption{$\mathcal{RA}^{11}_{\infty}$ and $\mathcal{R} \mathcal{A}_{\infty}^{21}$ 
 in a 2 vs. 1 reach-avoid game. \textbf{Top left figure:} We fix the position of a defender at $\boldsymbol{x}_D$ and visualize the set $\mathcal{RA}^{11}_{\infty}$. An attacker wins the game if its states are in the $\mathcal{RA}^{11}_{\infty}$ region. \textbf{Top right figure:} The positions of a defender and an attacker are fixed to visualize the slice $\mathcal{RA}^{21}_{\infty}$. At least one of the attackers can reach the target set if $\boldsymbol{x}_{A_1} $ lies inside the region $ \mathcal{RA}^{21}_{\infty} $. \textbf{Bottom figure}: We fix the positions of two attackers $\boldsymbol{x}_{A_1}, \boldsymbol{x}_{A_2}$ to visualize the slice $\mathcal{RA}^{21}_{\infty}$. The defender can capture both attackers if it lies outside the region $ \mathcal{RA}^{21}_{\infty}$.}
\label{2v1_RAS}
\end{figure}

First, we visualize the 4-dimensional 1 vs. 1 reach-avoid tube $\mathcal{RA}^{11}_{\infty}$ in a 2-dimensional plane by fixing the state of the defender at $(0.3, -0.3)$. If an attacker is within the reach-avoid tube, then it can win the game using optimal control inputs; otherwise, the defender wins. As shown in the top left subplot of Fig. \ref{2v1_RAS}, both attackers at $(-0.5, 0.0)$ and $(0.0, 0.8)$ lie out of the region; hence, each of these attackers will lose in a 1 vs. 1 game to the defender.
Similarly, to visualize the 6-dimensional reach-avoid tube $\mathcal{R} \mathcal{A}_{\infty}^{21}$, we fix the positions of attacker 1 at $(-0.5, 0.0)$ and the defender at $(0.3, -0.3)$, which are shown in the top right subplot of Fig. \ref{2v1_RAS}. The attacker at $(0.0, 0.8)$ lies out of the reach-avoid tube boundary, hence the defender at $(0.3, -0.3)$ could capture both attackers using the optimal control inputs. 

\subsection{The 2 vs. 1 Reach-Avoid Game Simulation}

\begin{figure}[H]
\centering
\includegraphics[width=1.0\linewidth]{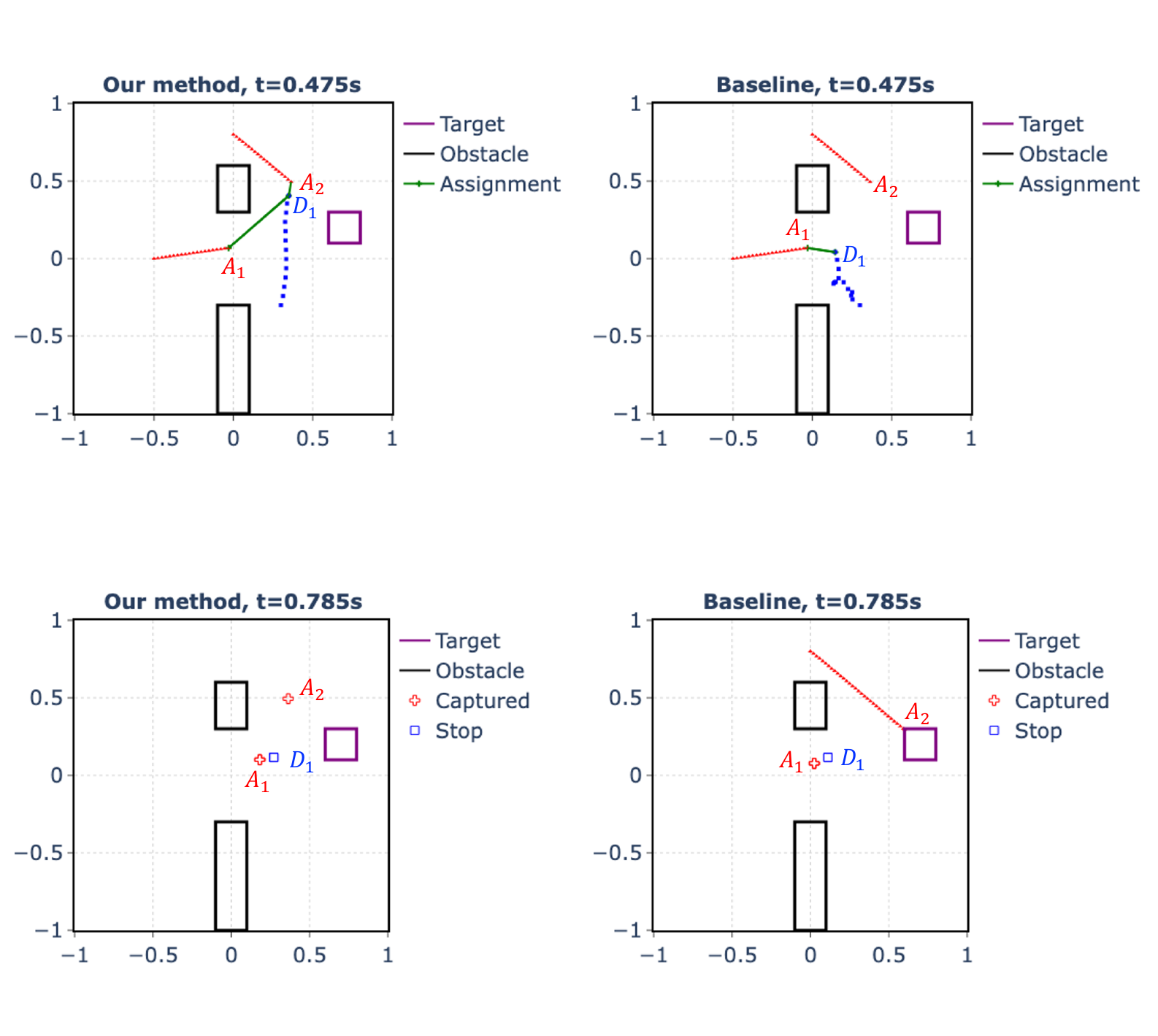}
\caption{A 2 vs. 1 reach-avoid game. In this figure, the plots on \textbf{left} column are simulations of the game using \textbf{our method} at different time steps. And the plots on the \textbf{right} are simulations using \textbf{baseline method} \cite{chen2014multiplayer}. At the beginning of the game, the defender is assigned to capture both attackers and successfully captures them at the end using our method. In contrast, using the baseline method, $A_{1}$ is captured first but then $A_{2}$ is able to reach the target. }
\label{2v1}
\end{figure}

Fig. \ref{2v1} shows the simulation results of two snapshots at $t=0.475s$ and $t=0.785s$. With our method, the defender was assigned to capture both attackers until attacker $A_2$ was captured at the time $t=0.475s$, and eventually, the defender can capture the attacker $A_1$.

As the left column in Fig. \ref{2v1} shows, at the beginning of the game, the defender at $(0.3, -0.3)$ is assigned to capture both attackers $A_1$ and $A_2$ by applying optimal control obtained by Eq. 
\ref{2v1d} using computed function $\Phi^{21}$. At the time $t=0.475s$, $A_2$ was captured, and at $t=0.690s$ $A_1$ was captured by the defender. Using our method, both attackers were successfully captured. 

In comparison, using the baseline method, the defender is only able to capture 1 attacker as shown in the right column of Fig. \ref{2v1}. Since any 1 vs. 1 assignment using the baseline is valid, initially the defender is assigned to capture attacker $A_1$ and applies the optimal control computed by Eq. \ref{1v1d} using $\Phi^{11}$, which resulted in $A_1$ being captured at the time $t=0.785s$. However, the resulting relative position of the defender $D_1$ allowed for attacker $A_2$ to reach the target set at the end of the game.

\subsection{The 6 vs. 2 Reach-Avoid Game Simulation}

\begin{figure}[H]
\centering
\includegraphics[width=1.0\linewidth]{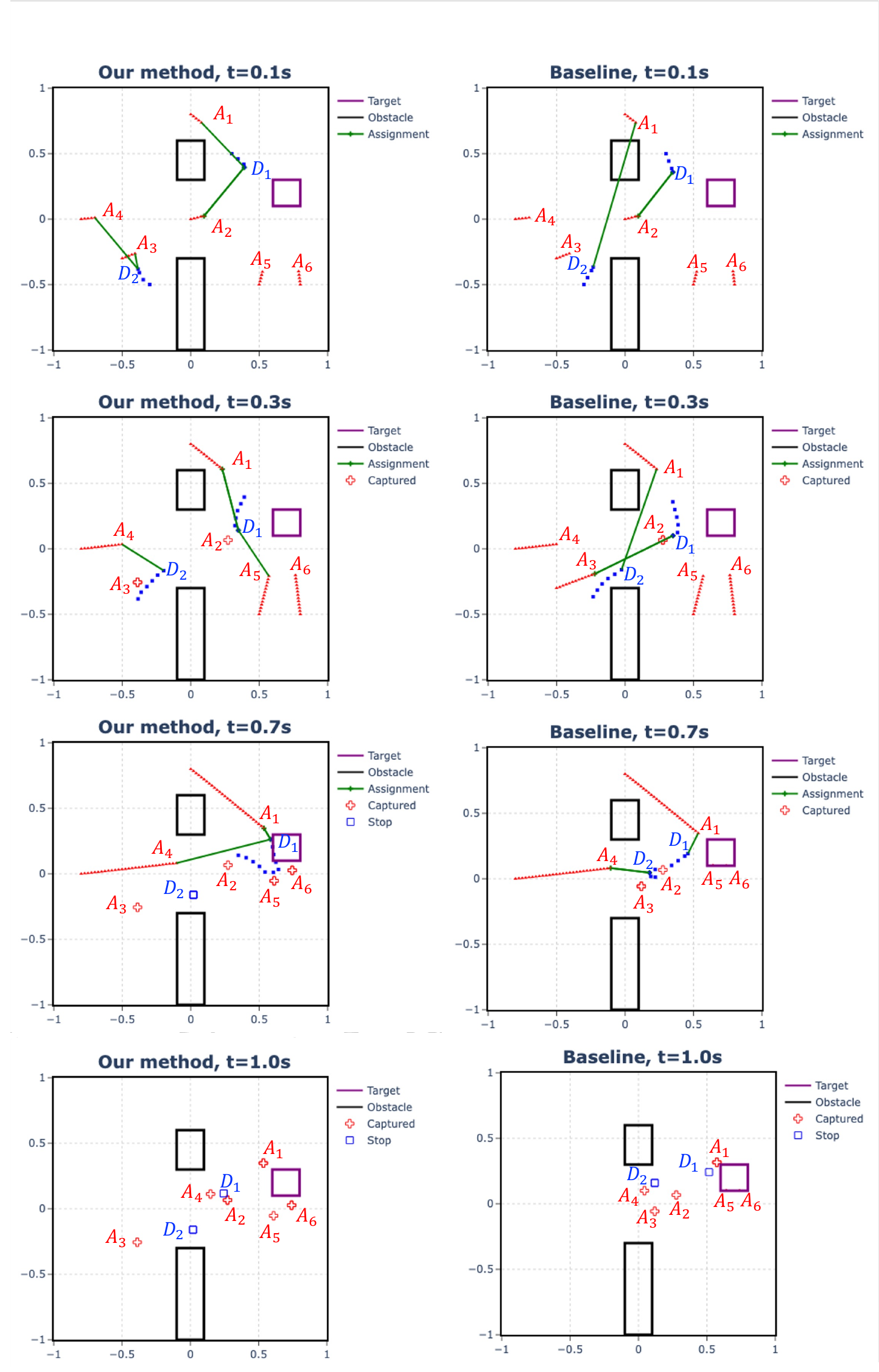}
\caption{6 vs. 2 game. In this figure, the plots on the \textbf{left} column are simulations of the game using \textbf{our method} at different time steps. And the plots on the \textbf{right} are simulations using \textbf{baseline method} \cite{chen2014multiplayer}. It can be observed at the end of the game, that our method can capture all attackers while the baseline can only capture 4 attackers with $A_5, A_6$ arrived at the target set.}
\label{6v2}
\end{figure}

Fig. \ref{6v2} shows different snapshots of a 6 vs. 2 game. Using our proposed method, 6 attackers were able to be captured. 

At the time $t=0.1s$, each defender was assigned to capture 2 attackers. After attacker $A_3$ was captured by defender $D_2$, the MIP assigned attacker $A_4$ only to $D_2$ as shown in the $t=0.3s$ snapshot while assigning $D_1$ to both $A_1$ and $A_5$. At the time $t=0.7s$, $D_1$ was assigned to capture the rest two attackers $A_1$ and $A_4$; at this time, $D_2$ did not get any more assignments and stopped. Finally, all attackers were captured at the time $t=1.0s$. The reduction of the total number of attackers over time is consistent with the aforementioned property of Algorithm \ref{alg1}.

In comparison, when using the baseline method, only 4 attackers were successfully captured eventually. 
During the whole game, attackers $A_5$ and $A_6$ did not match to any defender and arrived at the target eventually. With the baseline method, one attacker could be assigned to at most one defender at a time, which would result in the defender acting more shortsightedly and missing the chances of stopping more attackers.

\subsection{The 8 vs. 4 Reach-Avoid Game Simulation}

\begin{figure}[H]
\centering
\includegraphics[width=1.0\linewidth]{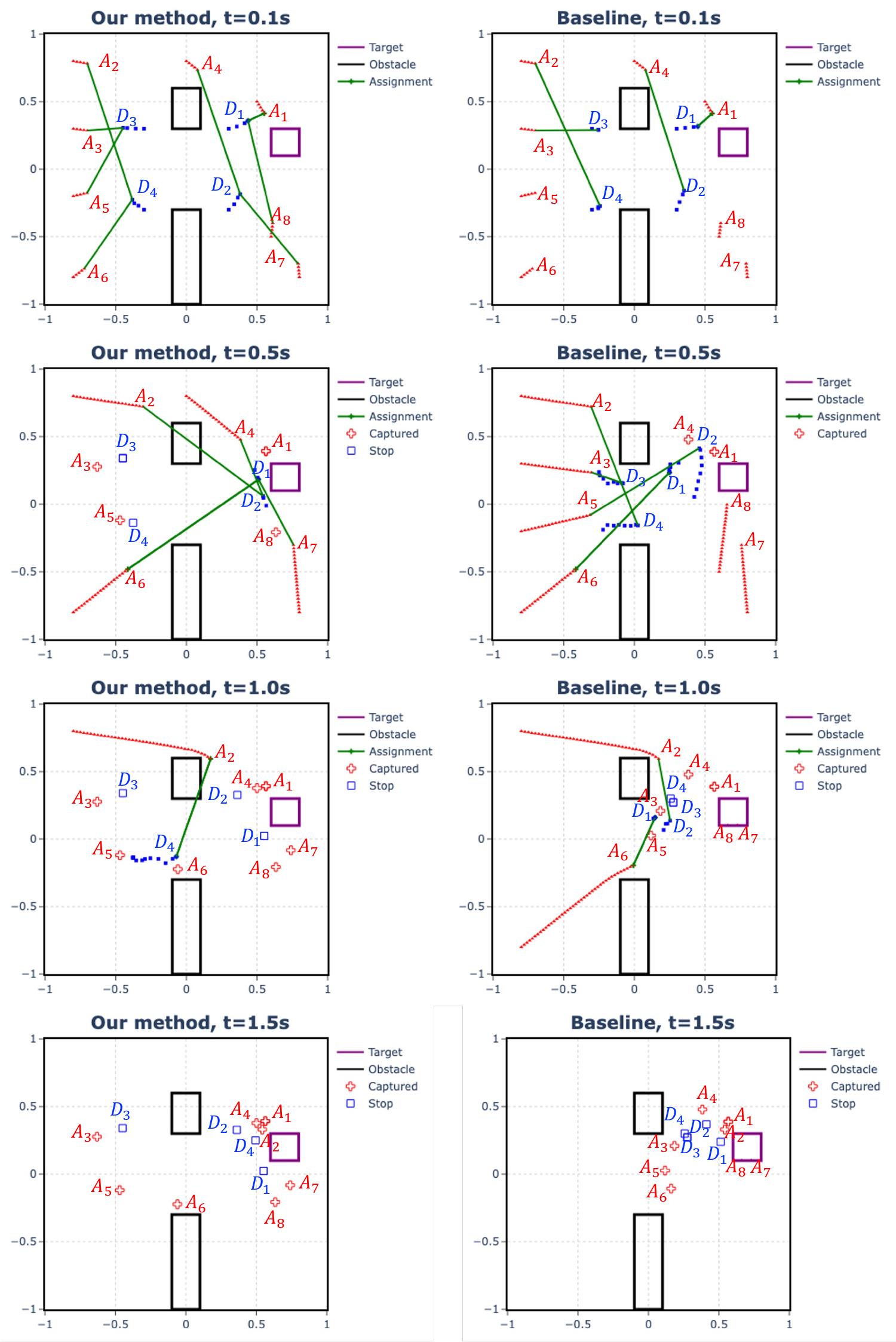}
\caption{ 8 vs. 4 game. In this figure, the plots on the \textbf{left} column are simulations of the game using \textbf{our method} at different time steps. And the plots on the \textbf{right} are simulations using \textbf{baseline method} \cite{chen2014multiplayer}. It can be observed at the end of the game, that our method can capture all attackers while the baseline can only capture 6 attackers with $A_7, A_8$ arrived at the target set.}
\label{8v4}
\end{figure}

Fig. \ref{8v4} shows the results of the 8 vs. 4 simulation. With our method, all 8 attackers were captured. Initially, at $t=0.1s$, each defender was assigned to 2 attackers, as shown in the left column of Fig. \ref{8v4}. Because of that, over the course of 0.5 seconds, 4 attackers $A_1$, $A_3$, $A_5$, and $A_8$ are already captured. At the time $t=1.0s$, there was only one attacker $A_2$ left, which is assigned to defender $D_4$. At the end of the game, all attackers fail to reach the target region without being captured. 

In comparison, using the baseline method, 6 attackers were captured, while attackers $A_7$ and $A_8$ were never assigned to any defenders, and hence reached the target set as shown in the \textbf{right} column Fig. \ref{8v4}. The number of matching assignment is at most 4 during the whole simulation which is the same as the number of defenders.

It can be observed from our simulation results that the baseline method would require additional defenders in order to capture all the attackers.

\section{CONCLUSION}
\label{sec5}

In this paper, we have proposed an HJ reachability-based method to compute the winning-guaranteed set of states for 1 defender in a reach-avoid game against 2 attackers utilizing recent computational advances. Furthermore, we approximate the solution to multi-agent reach-avoid games by proposing a MIP method that maximizes the number of captured attackers. We proved the effectiveness of our method in simulations and showed that our method works very well in cases where defenders are fewer than attackers. 

In the future, we will explore reach-avoid games with different configurations, for example, the 1 vs. 2 reach-avoid game with each agent 2-dimensional dynamics, and the 2 vs. 2 reach-avoid game where one of the agents has simple 1-dimensional dynamics. Besides, we will also use neural networks to approximate higher-dimensional games \cite{bansal2021deepreach}. The new optimization scheme that can efficiently assign matching for these games is also in the future plan.

\addtolength{\textheight}{-12cm}   

\bibliographystyle{ieeetr}
\bibliography{ref.bib}

\end{document}